\documentclass[10pt,emulateapj,apj]{emulateapj}
\shorttitle{Estimation of Gravitational Potential Shear field from Galaxy Redshift Survey}
\shortauthors{Park, Kim \& Park}
\begin{document}
\title{Gravitational Potential Environment of Galaxies: I. Simulation}
\author{Hyunbae Park\altaffilmark{1}, Juhan Kim\altaffilmark{2}, and Changbom Park\altaffilmark{3}}
\altaffiltext{1}{School of Physics and astronomy, Seoul National University, Gwanak-Gu,
Seoul 151-742, Korea; hcosmosb@snu.ac.kr}
\altaffiltext{2}{College of Applied Science, Kyung Hee University, Seocheon-dong,
Giheung-gu, Yongin-si, Gyunggi-do 446-702, Korea; kjhan0606@gmail.com}
\altaffiltext{3}{School of Physics, Korea Institute for Advanced Study, Heogiro 87, Dongdaemun-gu, 
Seoul 130-722, Korea; cbp@kias.re.kr}

\begin{abstract}
We extend the concept of galaxy environment from the local galaxy number density 
to the gravitational potential and its functions like the shear tensor. 
For this purpose we examine whether or not one can make an accurate estimation 
of the gravitational potential from an observational sample which is finite 
in volume, biased due to galaxy biasing, and subject to redshift space distortion. 
Dark halos in a $\Lambda$CDM simulation are used in this test.
We find that one needs to stay away from the sample boundaries by more than
30$h^{-1}$Mpc to reduce the error within 20\% of the root mean square values
of the potential or the shear tensor.
The error due to the galaxy biasing can be significantly reduced by using 
the galaxy mass density field instead of 
the galaxy number density field. The error caused by the redshift space
distortion can be effectively removed by correcting galaxy positions 
for the peculiar velocity effects.
We inspect the dependence of dark matter halo properties on four environmental
parameters; local density, gravitational potential, and the ellipticity
and prolateness of the shear tensor. 
We find the local density has the strongest correlation with
halo properties. This is evidence that the internal physical properties
of dark halos are mainly controlled by small-scale physics.
In high density regions dark halos are on average more massive and 
spherical, and have higher spin parameter and velocity dispersion.
In high density regions dark halos are on average more massive and
spherical, and have higher spin parameter and velocity dispersion.
We also study the relation between the environmental parameters and the subtypes of dark halos.
The spin parameter of satellite halos depends only weakly on the local density for 
all mass ranges studied while that of isloated or central halos depends more 
sensitively on the local density. 
The gravitational potential and the shear tensor have weaker
correlations with halo properties, but have environmental information
independent of the local density.

\end{abstract}


\section{Introduction}

One of recent developments in the study of galaxy formation is 
quantitative understanding of the dependence of galaxy 
properties on environment (Park et al. 2007; 
Hwang \& Park 2009 among others). It has already been noticed 
since the 1930's that galaxy luminosity and morphology depend 
on the local density: the high density regions preferentially 
harbor more luminous and early morphological type of galaxies 
(Hubble \& Humason 1931).
In the beginning of the studies on the environmental effects on 
galaxy properties (on the simulation side, see Lemson \& Kauffmann 1999; Gao et al. 2005; 
Jing et al. 2007; Hahn et al. 2007; Maccio et al. 2007; 
on the observation side, see Blanton \& Berlind 2007; Park et al. 2007; 
Cervantes-Sodi et al. 2008; Skibba et al. 2009; Blanton \& Moustakas 2009), 
the environment was distinguished according 
to the large-scale structure where the galaxies under study 
are located. For example, comparative studies for galaxies 
located within massive galaxy clusters, groups, or voids 
were carried out (Oemler 1974; Dressler 1980; Postman \& Geller 
1984; Rojas et al. 2004 among many others). 

Another trend of the same kind of study used continuous parameters 
that measure the local galaxy number density. Various kinds of 
smoothing kernel were used to estimate the local number density 
from galaxy positions in redshift space. The most popular one is 
the truncated cylindrical cone which is motivated from the fact 
that massive clusters appear as Fingers of God in redshift space 
(Hogg et al. 2004; Kauffmann et al. 2004; Kuehn \& Ryden 2005;
Reid \& Spergel 2009). 
The Gaussian and spline filters are also often used. 
These filters can have a fixed size or can vary in size to include 
a fixed number of galaxies (Park et al. 1994; Monaco \& Efstathiou 1999;
Park et al. 2007; Kim et al. 2008).

Even though these two approaches look quite different, they 
essentially use the local galaxy number density to distinguish 
among different environments. 
Extension of the concept of environment beyond the `local number density' 
has been started by several authors.
Park, Gott \& Choi (2008) and Park \& Choi (2009) used the 
mass density due to the galaxy plus dark halo systems
as a new environmental parameter. They also divided the `local'
density into the large-scale background mass density and
the small-scale density attributed to the nearest neighbor galaxy.
The mass density is estimated from galaxy luminosity 
and mass-to-light ratios. 
It turned out that the environment set up by the nearest neighbor 
was critically important in determining galaxy properties.
Lee et al. (2009) used galaxy luminosity density and local color 
(difference between luminosity densities in two bands) 
as environmental parameters.
Lee \& Lee (2008) inspected the relation between the ellipticity
of the tidal shear and galaxy morphology.

It is expected that some galaxy properties depend on the local 
galaxy number/mass density sensitively. Color and recent star 
formation activity may be such properties. However, the root-mean-square
(RMS) displacement  of mass is about $10 h^{-1}$Mpc till the present epoch 
in the $\Lambda$CDM model best fit to the Wilkinson 
Microwave Anisotropy Probe (WMAP) 3-year data (Park \& Kim 2009), 
and therefore the local density at the present location of galaxies 
cannot fully represent the environment where galaxies formed and 
evolved. This is true particularly for intermediate and high density 
regions since their sizes are typically only a few Mpcs. Therefore, 
it may be useful to consider environmental parameters other than 
local density to understand the environmental effects on galaxy formation.

A theoretically motivated environmental parameter is the gravitational 
potential. Dark matter and baryons are expected to fall into the deep 
gravitational potential well to form massive objects. 
Since the gravitational potential field picks up the fluctuation power 
at scales much larger than those of the density field, the correlation between them will 
not be perfect. And it will be interesting to see how galaxy 
properties are related with the `local' gravitational potential. 

Furthermore, it is expected that galaxy 
angular momentum is generated from the large-scale gravitational shear force. 
According to tidal torque theory 
(Hoyle 1951; Peebles 1969; Doroshkevich 1970;
White 1984; Lee \& Pen 2000; Vitvitska et al. 2002; Porciani et al 2002), 
the origin of galactic angular momentum is
originated by the
 tidal torque operating on primordial gas lump that will form a galaxy.
 The torque is given by $\tau = T \times I$,
 where $T$ is a shear tensor generated by external material
 and $I$ is a moment of inertial tensor of material being torqued.
 Navarro et al. (2004) presented supporting evidence for the theory
 with direction of galaxy rotation axis.
An accurate estimation of the tidal shear tensor will enable one
to verify if the tidal torque theory is really responsible for galaxy spin
(Porciani et al. 2002; Lee \& Pen 2002).

In this paper we will study how accurately one can estimate the gravitational
potential and its functions from a simulated sample of galaxies. 
The error sources in this estimation are divided into three categories:

1. finite volume of the survey, 

2. galaxy biasing, and

3. redshift space distortion.

\noindent
We then inspect the dependence of dark matter halo properties on
various environmental parameters including the `local' gravitational
potential. It is hoped that a generalization of environmental parameter 
beyond the local density allow us to better understand galaxy formation 
and evolution.


\section{Method}

\subsection{Simulation}

We will examine how accurately one can estimate the gravitational
potential and its derivatives when an observational sample is given.
For this purpose we use a set of mock galaxies identified as 
dark matter halos in an N-body simulation of the universe.
The simulation we use here adopted
the cosmological parameters measured from the 
WMAP 3-year data (Spergel et al. 2007), 
which are $\Omega_{\Lambda}=0.762, \Omega_m=0.238, \Omega_b=0.042,
n_s=0.958, h=0.732$, and $\sigma_8 = 0.761$, where $\Omega_{\Lambda},
\Omega_m, \Omega_b$ are density parameters due to cosmological constant,
matter, and baryon, respectively, $n_s$ is the slope of the primordial power spectrum,
and $\sigma_8$ is the RMS fluctuation of mass in an $8 h^{-1}$Mpc 
radius spherical top hat. 
It ran $2048^3$ Cold Dark Matter (CDM) particles 
whose initial conditions are generated  on a $2048^3$ mesh 
in accordance with the $\Lambda$CDM power spectrum.
The simulation was started at $z_i=47$ 
taking 1880 global time steps till the present epoch. 
The physical size of the simulation cube is $1024 h^{-1}$Mpc.
We used a parallel particle-mesh (PM) + tree N-body code (Dubinski et al. 2004) 
to increase the spatial dynamic range. 
The gravitational force softening length is set to 50$h^{-1}$kpc which is
0.1 times the mean separation between CDM particles.
The particle mass is $9.6\times h^{-1} 10^9 M_{\odot}$.

We identified the gravitationally bound, tidally stable dark matter halos 
(physically self-bound or PSB halos)
from the CDM particle data at $z=0$ (Kim \& Park 2006). At the first step
the Friend-of Friend (FoF) algorithm is used to search for dark halos
adopting the connection length of 0.2 times the mean particle separation. 
Then subhalos are identified within each FoF particle group 
taking into account the gravitational binding energy with respect
to the local density maximum and the tidal force from other more massive
subhalos if they exist. The minimum mass halos contain 30 member particles, and the 
halo mass function is accurate down to $M_h = 2.9 \times 10^{11} h^{-1}
M_{\odot}$. The mean separation between dark halos is $4.3 h^{-1}$Mpc.
The resulting dark matter halo sample consists of three types: 
isolated halo which does not
overlap with other halo (but it can have close neighbor halos),
central halo which is the most massive halo in each group of halos,
and satellite halo which is not the central one in each group of halos.

Galaxies are identified with the halos. This dark matter halo--galaxy
one-to-one correspondence model describes the observed galaxy 
distribution quite accurately (Kim et al. 2008; Gott et al. 2008; Gott et al. 2009).
Note that our dark halos are not the commonly used FoF halos to which 
the Halo Occupation Distribution prescription is usually applied to 
statistically distribute galaxies (Zheng et al. 2008 and references therein). 
Our dark halos can be central
galaxies or satellites, which are the direct results of the N-body simulation.
To match the dark halo sample with observed galaxy samples Kim et al. (2008)
adjusted the halo mass threshold, making the mean number densities of halos and
galaxies the same. The resulting halo sample was to be
compared with a volume-limited sample of galaxies brighter than an
absolute magnitude threshold.

\subsection{Potential Calculation}

In the simulation we know the positions of all CDM particles within 
the simulation cube with periodic boundaries. Therefore,
the gravitational potential can be calculated through the Poisson equation
\begin{equation} 
\Phi_{\bf k} = -{{4\pi G}\over k^2}{\bar\rho}\delta_{m{\bf k}} 
   = -{3\over 2}{{\Omega_m H^2}\over  k^2}\delta_{m{\bf k}},
\end{equation}
where ${\bar\rho}$ is the mean density, $\delta$ is the overdensity, and
we decomposed the gravitational potential and the density fields into Fourier modes.
This true gravitational potential is going to be compared with those obtained
by using galaxies (i.e. dark halos).

In practice, we cannot observe dark matter particles, but can observe
only galaxies.
We adopt a simple method to obtain the gravitational potential and its derivatives 
from a galaxy redshift sample.  We assume galaxies are locally biased tracers of 
the underlying mass: $\delta_g(x) = b \delta_m(x)$, where $b$ is the bias factor.
Then the Poisson equation in the Fourier space is
\begin{equation} 
\Phi_{\bf k} = -{3\over 2}{{\Omega_m H^2}\over b k^2}\delta_{g{\bf k}}.
\end{equation}
The proportionality factor between $\Phi_{\bf k}$ and $k^{-2}\delta_{g{\bf k}}$ is not
important in our study because each field will be normalized by its RMS value.
In section 3.2 we will compare the potential calculated from the galaxy
distribution with that from the dark matter distribution.

To use equations (1) and (2) we first calculate the density field from
matter particle or galaxies.  We assign them on a cubical mesh using the 
Triangular-Shaped Cloud algorithm (Hockney \& Eastwood 1981), and Fourier transform
the density array to get $\delta_{m{\bf k}}$ or $\delta_{g{\bf k}}$. 
The gravitational potential in real space is then obtained from equation (1) 
or (2), and is interpolated at each location of galaxies.
The gradient of $\Phi$ and the shear tensor are calculated by finite differencing
$\Phi$ in real space.  

Near a galaxy at ${\bf x}=0$ the gravitational potential 
can be given by the Taylor expansion
\begin{equation} 
\Phi({\bf x}) = \Phi(0)-\xi_i x_i - {1\over2}\lambda_i x_i^2,
\end{equation}
where $\xi_i$ are the components of local acceleration, and $\lambda_i$ are the 
eigen values of the shear tensor $T_{ij} = \partial_i\partial_j \Phi$.
The coordinate axes are assumed aligned with the principal axes of the shear tensor.
The trace of the shear tensor at the galaxy is
\begin{equation} 
\nabla^2 \Phi = \sum \lambda_i = 4\pi G (\rho(0)-{\bar\rho}).
\end{equation}
Bardeen et al. (1986) introduced the parameters 
\begin{equation} 
e = {{\lambda_1-\lambda_3}\over 2\sum\lambda_i}, p={{\lambda_1-2\lambda_2+\lambda_3}\over 2\sum\lambda_i},
\end{equation}
characterizing the ellipticity and prolateness of a (density) distribution.
Lee \& Lee (2008) adopted the $e$ parameter in their study on tidal shear
dependence of galaxy morphology. However, galaxies typically are not located 
at extrema of potential field unlike the density field at the smoothing scale
we will adopt ($R_G\sim 6h^{-1}$Mpc), and the sum of
eigenvalues can be zero, making the above parameters undefined.
This is particularly true for large-scale density and potential fields where
the galaxy-scale density peaks are not resolved.
To avoid this problem we adopt the asymmetry and prolateness parameters
\begin{equation} 
E = \lambda_1-\lambda_3, P=\lambda_1-2\lambda_2+\lambda_3
\end{equation}
normalized by their RMS values.
These parameters are correlated with the local density through Equation (4).
Dependence of galaxy properties on the shear tensor should be then studied 
at fixed local density.


\subsection{Peculiar velocity correction}

In section 3.3 we will study how the potential and its functions are
biased because they are estimated from the galaxy distribution in
redshift space. It is shown that most of the redshift space distortion
effects can be removed by making a linear estimation of the peculiar
velocity from the redshift space galaxy distribution. The linear regime
peculiar velocity is calculated as follows.
The linearized continuity equation is
\begin{equation} 
\nabla\cdot {\bf v} = -{{\partial\delta}\over{\partial t}}
=-{{\dot D}\over D}\delta,
\end{equation}
where ${\bf v}=d{\bf x}/dt$ is the peculiar velocity in comoving space, 
$\delta({\bf x},t)$ is the density contrast 
$\delta=(\rho-{\bar\rho})/{\bar\rho}$,
and $D(t)$ is the linear growth factor. If we define the velocity potential 
$\phi$ by $d{\bf x}/dD\equiv \nabla\phi$, we obtain 
\begin{equation} 
\nabla\cdot {\bf v}={\dot D}\nabla^2\phi.
\end{equation}
Combining the equations (7) and (8) yields
\begin{equation} 
\nabla^2\phi = -\delta/D.
\end{equation}
The peculiar velocity is then given by
\begin{equation} 
{d{\bf x}\over dt} ={da\over dt} {dD\over da}{{d{\bf x}}\over dD }
= D H f \nabla\phi,
\end{equation}
where $a(t)$ is the expansion parameter, 
$f = d\log D/d\log a$, and $H(t)={\dot a}/a$ is the Hubble parameter.
Substituting Equation (9) into Equation (10) in Fourier space, 
one finally obtains the relation between density and peculiar velocity
\begin{equation} 
({d{\bf x}\over dt})_{{\bf k}} = i H
f {{\bf k}\over k^2}\delta_{{\bf k}}.
\end{equation}
We actually use a biased tracer of matter field, namely galaxies.
If the galaxy number density fluctuation is related 
with the matter fluctuation by $\delta_g = b\delta$,
the factor $\delta_{{\bf k}}$ in Equation (11) is replaced
by  $\delta_{g,{\bf k}}/b$.

\section{Results}

\subsection{Boundary effects}

\begin{figure}
\includegraphics[scale=0.9]{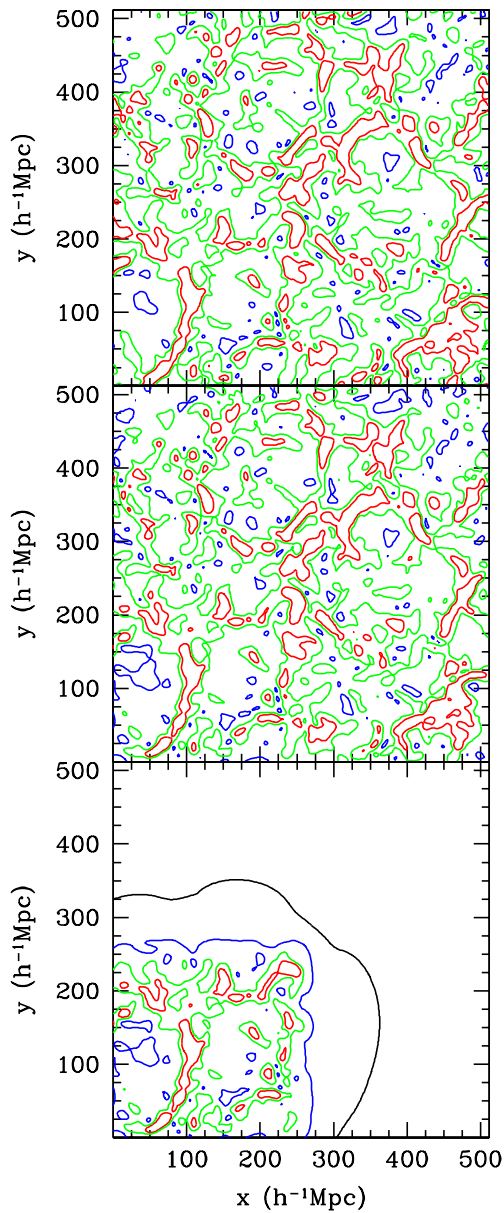}
\caption{Ellipticity of the gravitational shear tensor calculated from the whole 
simulation cube data of size 1024 $h^{-1}$Mpc (top), 
from a half-size subcube (middle),
and from a quarter-size subcube (bottom). Shown are constant shear magnitude
contours at four threshold levels in half-size slices. 
}
\end{figure} 

\begin{figure}
\includegraphics[scale=0.8]{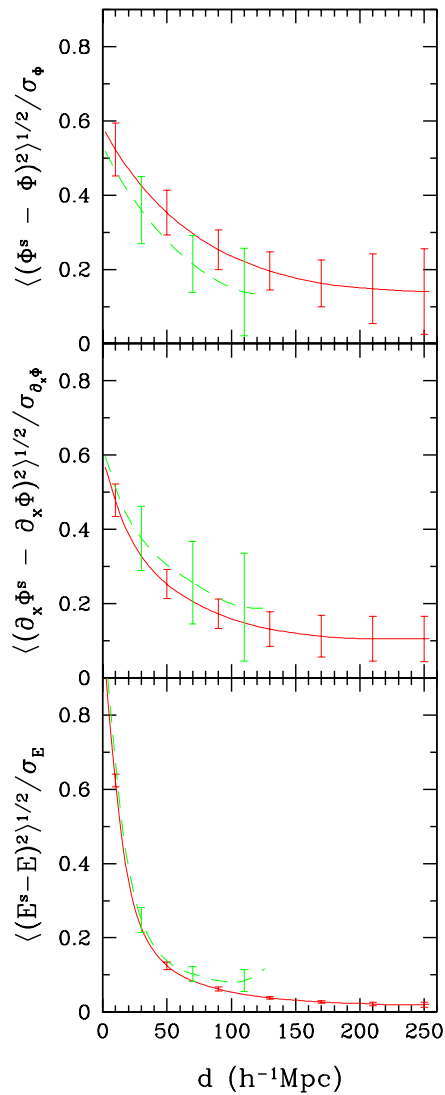}
\caption{(top) The Root-Mean-Square (RMS) error in potential estimated
from half-size (solid line) and quarter-size (dashed line) subcubes as a function
of distance from cube surfaces. The error is normalized with respect to
the RMS variation of potential. (middle) Same but for the $x$-component of
the potential gradient. (bottom) Same but for the ellipticity of the shear tensor.
}
\end{figure}

The gravitational potential would be correctly calculated if the mass distribution
were given over the infinite space. In practice, any observational sample covers
a finite volume of the universe with complicated boundaries.
In this section we will study the effects of the missing data beyond
the sample boundaries on gravitational potential and its derivatives.

We first calculate the reference potential from all dark halos in the whole
simulation cube. Since we are interested in large-scale environment, we smooth the 
density field by a $6h^{-1}$Mpc Gaussian. This scale is also motivated by the 
fact that the mean separation between $M_*$ galaxies is about  $6h^{-1}$Mpc
(cf. the `Best' sample of Park et al. 2005; see also Choi et al. 2007).
We then calculate a potential using only the dark
halos in the half-size cube near a corner of the simulation box. The remaining part
of the simulation cube is replaced by the mean dark halo number density estimated from
the half-size cube. To inspect the boundary effects further, we also calculate a potential
from the dark halos in the quarter-size cube at the same corner.

In Figure 1 we compare the three gravitational potentials in a slice passing
through a $z$-coordinate of $z=128 h^{-1}$Mpc of  the simulation cube. 
The slice includes the center of the quarter cube.  The top panel shows 
the shear ellipticity field $E/\sigma_E$ in the slice calculated from all
dark halos in the simulation. We show only the bottom left corner of the slice
where a comparison with the half cube result is possible. The contour levels
correspond to $1\sigma$ high (red contours), mean, $1.5\sigma$ low,
and $2.5\sigma$ low (black) shear ellipticity.
In the middle panel the shear ellipticity field calculated from the dark
halos in the half cube is shown. The contour levels are the same.
It is clear that the shear field of the half cube is extraordinarily similar 
with that of the full cube. This is particularly true near the center of
the half cube, namely at the position (256, 256). However, large differences 
are also observed near the boundaries. 
The bottom panel shows the shear field of the quarter cube.
It again shows that the shear ellipticity of the quarter cube agree quite well
with those of the other ones, particularly near the center of the quarter cube. 
It can be seen that there is a weak shear field extending 
beyond the sample boundary. The shear field of the half cube also has such leakage
outside the boundaries (truncated in the middle panel).

To quantitatively estimate the accuracy of the potential and its derivatives 
obtained from subcubes we calculate the RMS differences of the fields within 
a cubical shell centered on the center of the subcubes. 
The solid lines in the top panel of Figure 2 are the difference between
the potential from the full cube and that of the half cube within 2 $h^{-1}$Mpc-thick 
cubical shell centered at $(x, y, z)=(256 h^{-1}{\rm Mpc}, 256 h^{-1}{\rm Mpc}, 
256 h^{-1}{\rm Mpc})$. 
Size of a cubical shell is  2($257h^{-1}{\rm Mpc}-d$), where $d$ is
the distance of the shell from the surface of the subcube.
For example, $d=1h^{-1}$Mpc corresponds the outermost shell of 512$h^{-1}$Mpc size.
The dashed line is for the quarter cube.
The top panel shows how the error in the potential varies as one moves
into the subcubes. Lines and error bars are the mean and RMS differences
obtained from eight subcube results.
The difference monotonically decreases as the distance from boundaries increases.
Since potentials can be added by a constant without altering physics,
the difference of about 0.14$\sigma_{\Phi}$ near the center can be ignored
in the case of half cube. The figure tells that, if the potential field is to
be obtained with error less than say 20\% of its RMS value, one should stay 
more than $54h^{-1}$Mpc from all sample boundaries.
The quarter cube results show a much larger variance that the half cube results.

A similar trend is seen for the potential gradient as shown in the middle panel of Figure 2.
The $x$-component of the potential gradient mimics the linear radial peculiar velocity field
very far from an observer.
Due to lack of data outside subcubes the error in gradient is large near boundaries
but monotonically decreases as $d$ increases. The difference is smaller if one
ignores the large-scale gradient on the scale of the subcube.

The bottom panel of Figure 2 compares among the shear ellipticity 
from the full, half, and quarter cubes. 
It is important to note
that the error drops as the distance from boundaries increases
essentially in the same way for the half and quarter cube cases. The error 
reaches 20\% of the RMS shear ellipticity at the distance of $d=33 h^{-1}$Mpc.
The depth of this buffer region must be a function of the amplitude and
shape of the power spectrum.
We conclude that one needs to have an observational sample much larger than
60$h^{-1}$Mpc to secure the regions where the shear field has error less than
20\% of its RMS fluctuation.



\subsection{Effects of Galaxy Biasing}

\begin{figure}
\epsscale{0.9}
\plotone{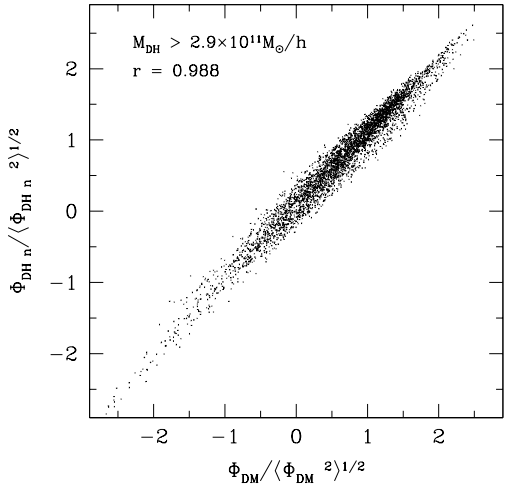}
\caption{Correlation between the potential estimated from the dark halo
number density ($\Phi_{\rm DH n}$) and that from dark matter distribution
 ($\Phi_{\rm DM}$).
Dark halos with mass larger than $2.9\times 10^{11} h^{-1} M_{\odot}$ are used
in the potential calculation.}
\end{figure}

\begin{figure}
\includegraphics[scale=0.8]{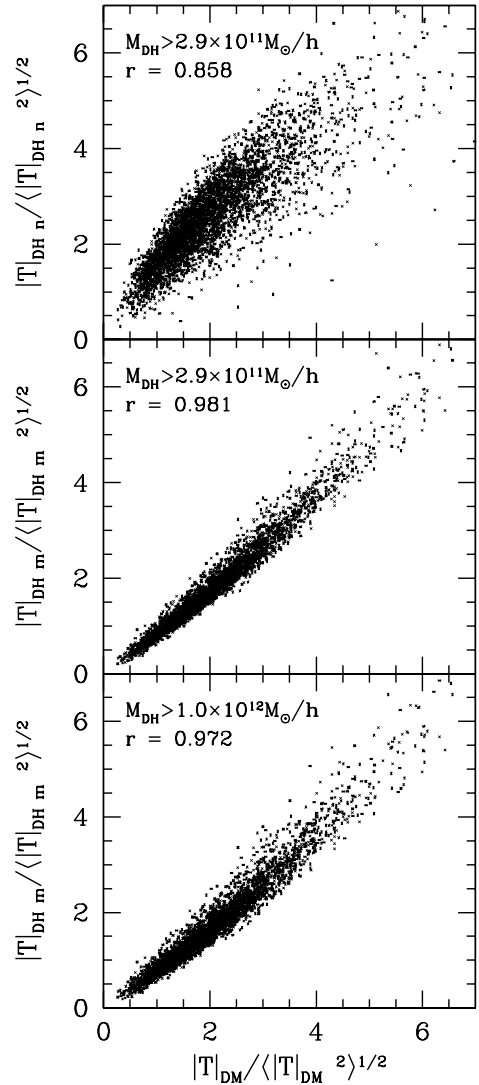}
\caption{
Correlation between the shear tensor magnitude calculated from the dark
matter distribution ($x$-axis) and those from dark halo number (top)
and dark halo mass (middle and bottom) distributions.
Dark halos with mass larger than $2.9\times 10^{11} h^{-1} M_{\odot}$ are used
in the top and middle panels. The mass cut is raised to  $1\times 10^{12} h^{-1} M_{\odot}$
in the bottom panel.
}
\end{figure}

\begin{figure}
\epsscale{0.82}
\plotone{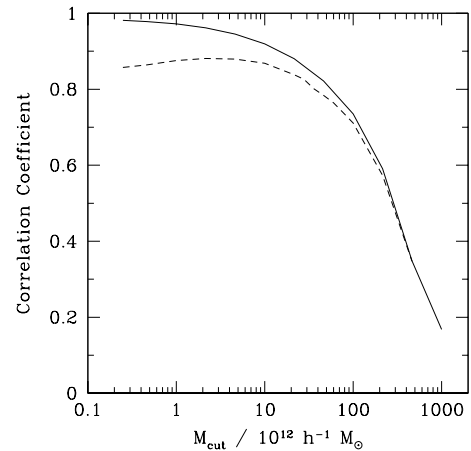}
\caption{
Correlation coefficient between the shear tensor magnitude calculated from the dark
matter distribution and those from dark halo number (dashed line)
and dark halo mass (solid line) distributions as a function of halo mass cut.
}
\end{figure}

In the previous section we used the number density of dark halos to calculate 
the potential field as it is common to use the galaxy number 
density to define environment of galaxies. 
However, it is the matter field that determines the true gravitational 
potential field, and one needs to understand the relation between the potential 
from mass and those from mass tracers.  In this section we assume that each 
and every PSB dark halo contains one galaxy, and compare the potential calculated 
from their number density or halo mass density with that from CDM particles 
in the simulation.

In Figure 3 we compare the potential from galaxy number density ($y$-axis)
with the correct potential from CDM particles. Dark halos more massive than 
$2.9\times 10^{11}M_{\odot}/h$ are used for the number density calculation.  
Their mean separation is 
$4.6 h^{-1}$Mpc, which is equal to that of the SDSS galaxies brighter 
than $M_r=-19.5+5{\rm log}h$ (Choi et al. 2007).
The correlation coefficient between the two potential fields is fairly high 
($r=0.988$).  At $\Phi_{DM}=0$ the potential from the galaxy number density 
has a dispersion of 0.20 times the RMS potential value.

Figure 4 compares between the magnitudes of the traceless shear tensor
$T_{ij} = \partial_i\partial_j \Phi - {1\over 3}\delta^K_{ij}\nabla^2\Phi$,
estimated from matter density, halo number density, and halo mass density fields.
Top panel shows that the correspondence between the shear fields from 
the galaxy number density and matter density is not so good. 
The correlation coefficient is only 0.858, and the dispersion 
of the shear magnitude obtained from galaxy number density is 0.47 times
the RMS shear magnitude at $|T|_{DM}/\langle|T|^2_{DM}\rangle^{1/2}=2$. 

Accuracy in potential and shear can be greatly improved if the halo mass is
used to weigh galaxies and the halo mass density field, instead of the number
density field, 
is used to calculate the potential. The second and third panels of Figure 4 
demonstrate such improvement when the halo mass threshold is set to 
$2.9\times 10^{11}M_{\odot}/h$ and $1.0\times 10^{12}M_{\odot}/h$, 
respectively. The latter objects corresponds to the SDSS galaxies 
brighter than $M_r\approx-20.4+5{\rm log}h$, 
close to that of the $M_*$ galaxies (Choi et al. 2007), in the sense that 
their mean separations ($6.6 h^{-1}$Mpc) are the same.
In the middle panel, at $|T|_{DM}/\langle|T|_{DM}\rangle^{1/2}=2$, 
the dispersion in $|T|_{DH}$ is only 0.17 times the RMS shear magnitude, 
an improvement by almost a factor 3. 
The correlation drops as the halo mass threshold
increases, but it still remains quite good for massive dark halos 
with $M>1.0\times 10^{12}M_{\odot}/h$.

Figure 5 shows how the correlation changes as a function of the halo mass
threshold. The solid line is the case when the halo mass density field is
used, and the dashed line is for the halo number density. The correlation
drops rapidly as the threshold increases above $10^{13}h^{-1}$M$_{\odot}$.
We conclude that the gravitational potential can be estimated quite accurately
by using the observed galaxy distribution, but accuracy can be greatly 
improved if the total mass associated with the dark halo plus galaxy system
is used to weigh galaxies in the shear tensor calculation.

When one chooses to use the mass field instead of the number density field,
one needs to adopt a halo mass estimator.
We suggest to use the red-band optical luminosity together with the
morphology-dependent mass-to-light ratios to estimate the relative mass of galaxies.
An example of using $r$-band luminosity as the mass estimator for dark halo plus galaxy
systems, can be found in Park \& Choi (2009), where the method turned out to
work quite well in the sense that galaxy properties show interesting
dependence on local and global environments at physically meaningful scales.
Note that we don't need to know the absolute value of halo mass if
parameters are normalized by their RMS values as we do here.
Using any halo mass estimator monotonically proportional to the actual halo mass
will improve the accuracy of the resulting potential field.

\subsection{Redshift-space Distortion}

\begin{figure}
\epsscale{0.9}
\plotone{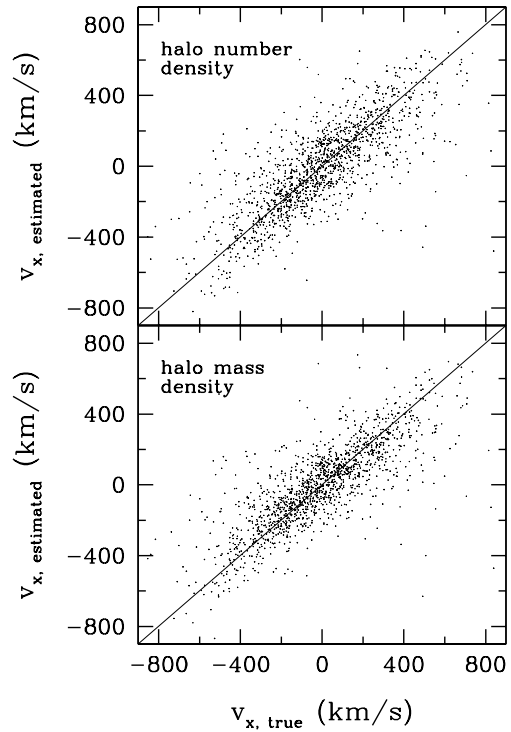}
\caption{The $x$-component of peculiar velocity estimated from halo number density (top
panel) and from halo mass density (bottom) compared with that from matter density.
}
\end{figure}  

\begin{figure}
\includegraphics[scale=0.9]{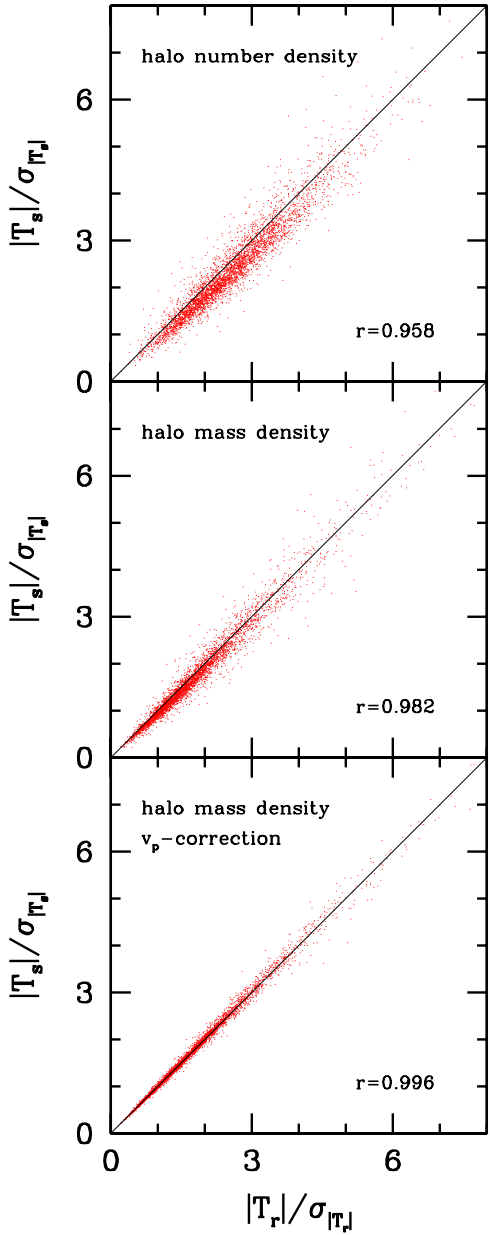}
\caption{(top) Magnitude of shear tensor calculated from 
redshift space halo number density versus that from matter distribution. 
(middle) Same, but for redshift space halo mass density.
(bottom) Same, but for redshift space halo mass density with the linear correction
for the redshift effects.
The numbers at the lower right corner are correlation coefficients.
}
\end{figure}  

When galaxy distance is obtained from redshift, the galaxy distribution becomes
biased in such a way that clusters and groups are stretched, filaments appear
more prominent by broadening the interior but compressing the exterior, 
and voids look elongated along the line of sight (Kaiser 1987).
These redshift space distortion effects generate error when the gravitational
potential is estimated directly from the galaxy distribution in redshift space.
We will show here that this error can be almost entirely removed by making a linear
correction to the redshift space distribution of galaxies.

We calculate the peculiar velocities of dark halos from the dark halo number
density or mass density in redshift space using Equation (11). They are
compared with the true peculiar velocities in Figure 6.  The upper panel shows 
the $x$-component of the peculiar velocity at each galaxy location calculated
from halo number density, and the bottom one shows that from halo mass density. 
It can be seen that the relation is a little tighter in the second case.

The top panel of Figure 7 shows the correlation between the shear magnitudes 
estimated from halo number density and from matter density. The correlation 
becomes stronger when the halo mass density is used for the reconstruction
as shown in the middle panel. The redshift space distortion effects become
tiny when the redshift space positions of dark halos are corrected for
the peculiar velocity, which is calculated from the redshift space distribution
of dark halos. This is shown in the bottom panel.
Ideally one can iterate this correction process, but we found that one time
correction was sufficient.

\subsection{Environmental Dependence of Halo Properties}

\begin{figure}
\includegraphics[scale=0.7]{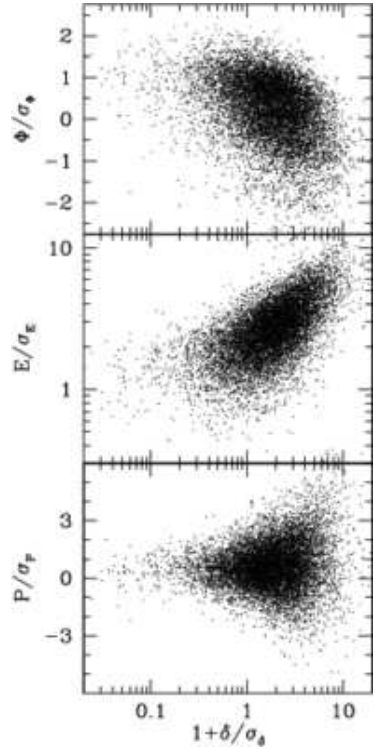}
\caption{Correlation of local density with gravitational potential (top),
ellipticity (middle), and prolateness (bottom) of shear tensor.
All are estimated from the dark matter density field smoothed over 6$h^{1}$Mpc.
}
\end{figure}  

\begin{figure*}
\includegraphics[scale=0.9]{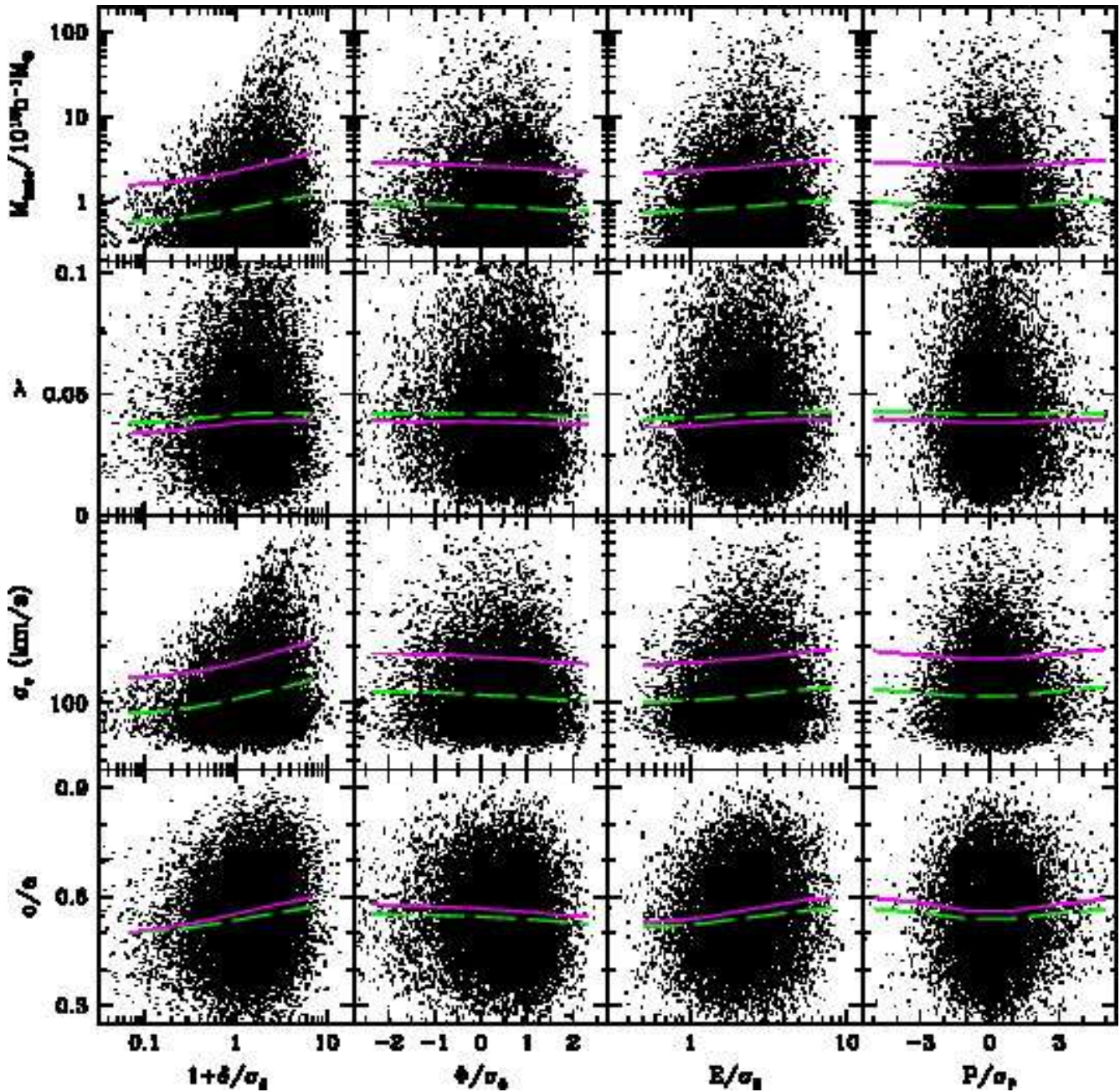}
\caption{
Relations between physical properties of dark halos and environmental parameters
for all types of halos with mass above $M_h > 2.9\times 10^{11} h^{-1}$ M$_{\odot}$.
Considered halo properties are mass (top row), spin parameter (second row), velocity
dispersion (third row), and short-to-long axis ratio (bottom row).
The left column shows the relation as a function of overdensity in the dark
matter density field smoothed over 6$h^{-1}$Mpc. In the second, third and fourth columns 
gravitational potential, ellipticity and prolateness of the shear tensor are used as
environmental parameters, respectively.
Dots are halos with mass  $M_h > 2.9\times 10^{11} h^{-1}$ M$_{\odot}$, and 
the solid lines are the mean relations. Dashed lines are for halos with 
$M_h > 1.0\times 10^{12} h^{-1}$ M$_{\odot}$.
}
\end{figure*}  

\begin{figure*}
\includegraphics[scale=0.85]{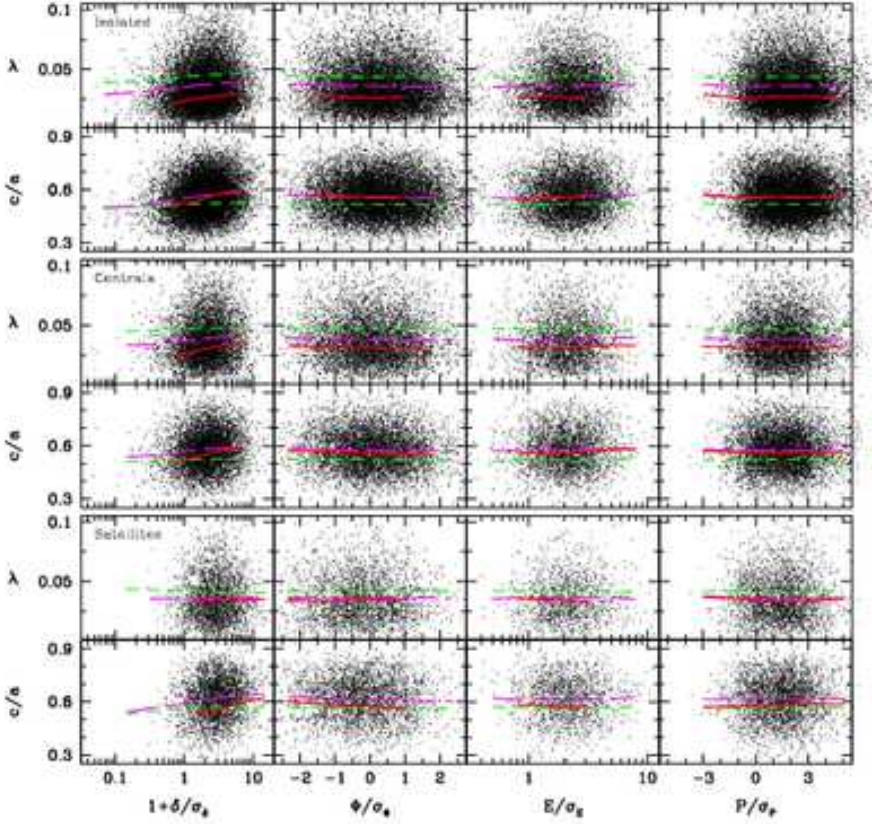}
\caption{
The spin parameter $\lambda$ and axis ratio $c/a$
for isolated halos (top two rows), central halos (middle two rows), and
satellite halos (bottom two rows) as functions of environmental parameters
used in Figure 9.
Dots in each panel are halos with mass $3\times 10^{12}h^{-1}
{\rm M_{\odot}} < M_h < 5\times 10^{12}h^{-1} {\rm M_{\odot}}$ and the long-dashed
lines are their mean relations. The short-dashed lines are for halos with
$3\times 10^{11}h^{-1} {\rm M_{\odot}} < M_h < 5\times 10^{11}h^{-1} 
{\rm M_{\odot}}$, and solid lines are for halos with
$3\times 10^{13}h^{-1} {\rm M_{\odot}} < M_h < 5\times 10^{13}h^{-1} 
{\rm M_{\odot}}$.
}
\end{figure*}  

\begin{figure*}
\includegraphics[scale=0.9]{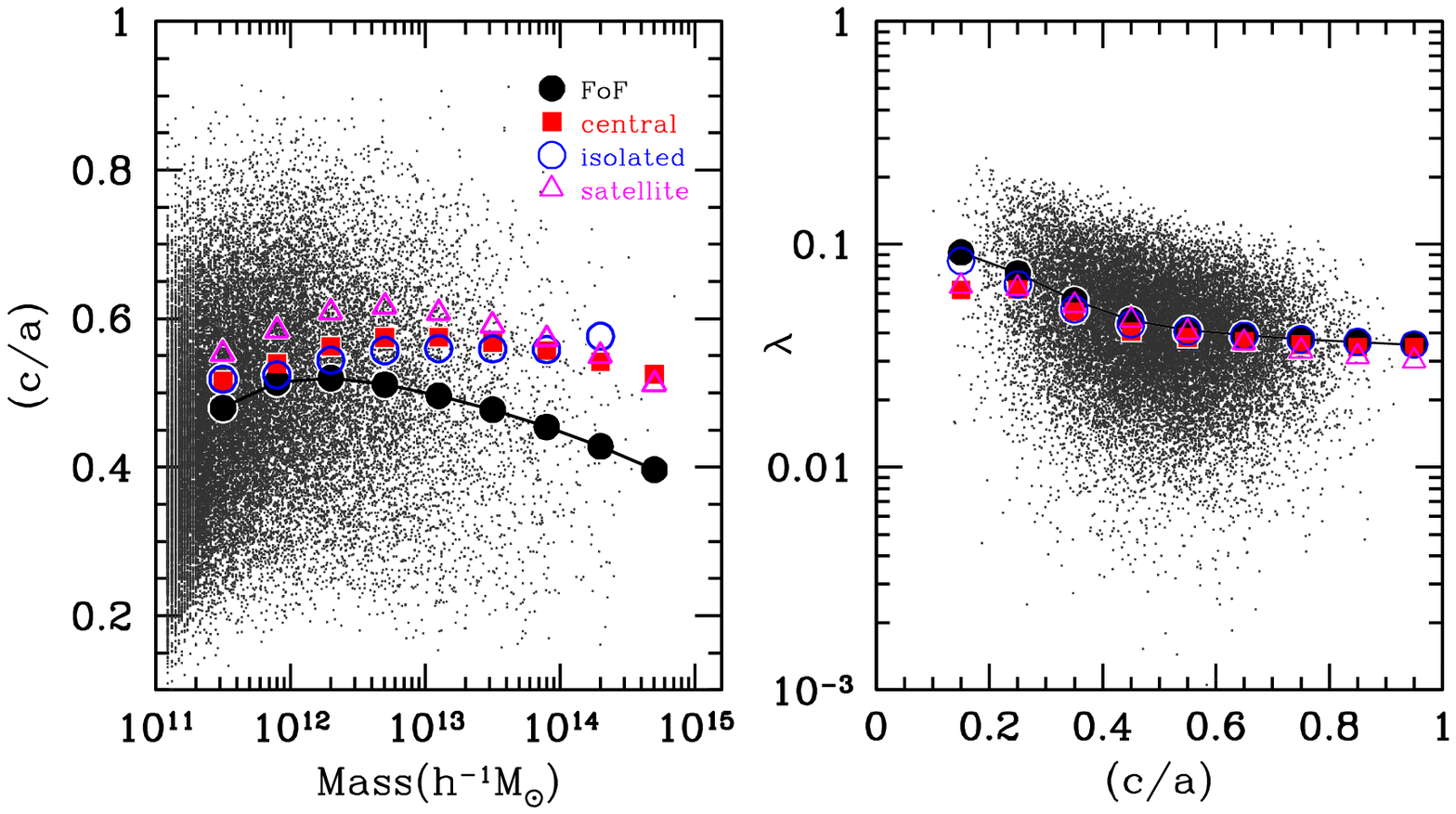}
\caption{Dependence of $c/a$ on the halo mass ({\it left})
and on the spin parameter ({\it right}).
The filled circles, filled boxes, open circles, and the open triangles
show the mean distributions of the FoF, central, isolated, and satellite halos.
Gray dots show the scatter plots of 30,000 halos selected in the FoF sample.
}
\end{figure*}

In previous sections we studied how accurately one can estimate the gravitational
potential and its functions when only biased and finite samples are available.
Now we will examine how various physical properties of dark matter halos
depend on environment including gravitational potential or its functions. 
The purpose is to see if their
dependence on gravitational potential or its functions is different from that on
local density. So it is necessary to see the correlation between
gravitational potential and local density at galaxy positions.
Figure 8 shows the potential (top), ellipticity (middle),
and prolateness of the shear tensor (bottom panel) as a function of 
overdensity. Each variable is normalized by its standard deviation.
It demonstrates that, even though there exists an overall correlation 
with the local density, the dispersion of the potential or shear tensor 
at fixed local density is very large.
This reflects the fact that the gravitational potential 
picks up large scale power (see Eq. 1), and thus
has information on environment independent of local density.

In Figure 9 we shows four physical parameters of dark matter halos as a function of 
local density, potential, ellipticity, and prolateness of the shear tensor.
Points are the dark halos with mass greater than $2.9\times 10^{11}h^{-1}$M$_{\odot}$.
Dashed and solid lines are the mean relations for dark halos with 
 $M_h > 2.9\times 10^{11}$ and $1\times 10^{12}h^{-1}$M$_{\odot}$, respectively.
Only 1/800 of dark halos in the simulation are plotted, but the mean relations are
obtained from 1/8 of dark halos. 
As expected, massive halos form in high density regions (top-left panel) because of
high interaction and merger rates (Park, Gott, \& Choi 2008; Fakhouri \& Ma 2009).
This plot very much resembles the local density dependence of galaxy luminosity obtained from
the CfA survey (Fig. 12 of Park et al. 1994) and the SDSS survey (Fig. 8 of Park et al. 2007).
It also tells that the maximum mass which a dark halo can acquire 
is a function of the local
density. In low density regions massive halos simply cannot form by the present epoch.
A similar observation can be made from the panel showing the velocity dispersion
of dark halos (the left panel of the third row).
The mean halo mass is relatively higher in low potential regions or 
in high shear ellipticity and prolateness regions,
but its dependence on these parameters is weaker than that on local density.
This dependence of halo mass on potential and shear is partly due to
the correlation of potential and shear with local density.

The second row of Figure 9 shows dependence of the spin parameter $\lambda$
on the environmental parameters. The spin parameter is defined by 
(Peebles 1969, Gardner 2001, and for an alternative definition see,
Bullock et al. 2001, Shaw et al. 2006, Hetznecker \& Burkert 2006)
\begin{equation} 
\lambda = J_{\rm vir} |E_{\rm vir}|^{1/2} / G M_{\rm vir}^{5/2},
\end{equation}
where $M_{\rm vir}, J_{\rm vir}$ and $E_{\rm vir}$ are the mass, total angular 
momentum and energy of a dark halo, respectively.
Contrary to other studies (Maccio et al. 2007) we detect dependence of $\lambda$
on all four environmental parameters.
For halos of $M_h > 10^{12}h^{-1}$M$_{\odot}$ the spin parameter is about 0.034 
in very low density regions, but increases to about 0.039 in high density regions.
Dark halos also tend to have higher spin in high shear magnitude regions. 
However, it should be noted that dark halo spin is not sensitive
to any of these environmental parameters. In particular, dependence of the spin 
on the tidal shear is weak, showing only about 4\% difference as the 
environment changes from low to high shear ellipticity regions (solid line in the
third panel of the second row of Figure 9).

The panels at the bottom of Figure 9 show that dark halos are more spherical
in high density, low gravitational potential, or high shear regions.
Such a trend is stronger for more massive halos. On average, high-mass halos
are preferentially more spherical than low-mass ones, and the difference 
is larger in high density/shear regions. This is consistent with the 
observational finding of Park et al. (2007) 
that early-type galaxies in the SDSS survey are rounder
as their luminosity increases (see their Figure 15). However, it
contrasts with an opposite finding in simulations by
many authors (Avila-Reese et al. 2005; Allgood et al. 2006; 
Maccio et al. 2007, 2008) 
who reported that low-mass halos are more spherical
than high-mass halos at a given epoch in their simulations.
The main reason for this discrepancy is the difference in the
halo definition as shown below.

To study the dependence of halo properties on environmental parameters 
in a more sensible way we divide our halo sample into subsets according to
halo mass and type. We consider three halo types: isolated,
central, and satellite halos.
Since halos can have a different growth history depending on their
type, it will be interesting to see the halo property-environment relation 
separately. We are also going to fix the halo mass
to subtract the halo mass dependence of a parameter from its environmental
dependence.  Figure 10 shows the spin parameter and axis ratio 
for isolated (top two rows), central (middle two rows), and
satellite (bottom two rows) halos. 
Dots in the top two rows are the isolated halos with $3\times 10^{12}h^{-1}
{\rm M_{\odot}} < M_h < 5\times 10^{12}h^{-1} {\rm M_{\odot}}$ and the long-dashed
lines are the mean relations. Only 1/25 of halos are plotted.
The short-dashed lines are for the halos with 
$3\times 10^{11}h^{-1} {\rm M_{\odot}} < M_h < 5\times 10^{11}h^{-1} 
{\rm M_{\odot}}$, and solid lines are for the halos with
$3\times 10^{13}h^{-1} {\rm M_{\odot}} < M_h < 5\times 10^{13}h^{-1} 
{\rm M_{\odot}}$.
We plot only the spin and $c/a$ parameters here because
the velocity dispersion shows almost no dependence on environment
in these subsamples with fixed halo mass ranges,
where $c/a$ is the ratio of the minor and major radii in a triaxial 
shape (Chandrasekhar 1969; de Zeeuw \& Franx 1991; 
Jing \& Suto 2002; Smith \& Watts 2005; Allgood et al. 2006).
In the middle and the bottom two rows of Figure 10 the points and 
lines correspond to the same mass ranges as in the top two rows 
but they are for central and satellite halos, respectively.

A few observations can be made from Figure 10. First, the dependence
of $\lambda$ and $c/a$ on the potential and shear is still very weak
(see the second, third, and fourth columns of Figure 10).
When the parameters are studied as a function of local density,
we find diverse relations that depend not only on halo mass but also
on halo type. 
An interesting finding is that the spin parameter of satellite halos tend to 
be constant or even decrease as the local density increases 
while that of isolated or central halos increases. 
It seems that the spin of satellites has decreased as they
interact more frequently with other halos including the central
halo in higher density environment. 
An observational evidence for this interpretation is that the spin of
late-type galaxies decreases as they approach their neighbors within
the virial radius (Cervantes-Sodi et al. 2010).
There is a clear trend for the spin parameter to decrease as the halo
mass increases for all halo types. The spin and local density 
tend to be more positively correlated for more massive halos.

It is also found in Figure 10 that halos are more spherical 
in higher density regions for all three types.
Furthermore, the dependence of $c/a$ on local density is stronger for more 
massive halos. Following the same interaction picture, we interpret this 
phenomenon as a result of tidal interactions between neighboring halos.
Halos in high density regions will suffer from the strong tidal force of 
neighboring halos more frequently, and can become more spherical 
on average. 
Our results are consistent with Maccio et al. (2007) who
found isolated and central halos are more spherical in high
density regions when the density is smoothed by an $8h^{-1}$Mpc tophat
even though their main conclusion was independence of halo properties
on large-scale background density.

The dependence of halo shape on halo mass is not monotonic.
Figure 10 shows halos are more spherical as mass increases from
$\sim 4\times 10^{11}{\rm M_{\odot}}$ to 
$\sim 4\times 10^{12}{\rm M_{\odot}}$, then more elongated as mass
increases to $\sim 4\times 10^{13}{\rm M_{\odot}}$. 
This result might seem dissonant with some previous reports that more massive halos 
are less spherical (Avila-Reese et al. 2005; Allgood et al. 2006; Maccio et al. 2007).
The disagreement may originate from difference in halo type and 
mass range. Avila-Reese et al. (2005) and Maccio et al. (2007) used the FoF or
spherical overdensity (SO) algorithm to identify halos, respectively. Therefore,
a halo in these works can be a group of halos connected or embedded with 
one another and can have multiple density peaks. The massive halos located at 
the intersections of dense filaments are more likely to be connected with 
other halos along the filaments by tenuous bridges of particles 
and appear more elongated.  On the other hand, the halos in our work 
(the PSB halos) have one and the only one density maximum, and are separated 
into isolated, central, and satellite halos depending on their positions
and mass. The PSB halos are to be directly identified with galaxies individually
(Kim et al. 2008). Therefore, the shape of the PSB halos does not necessarily
the same as the FoF or SO halos as the halo shape depends on halo 
definition (Bett et al. 2007).

The correspondence of our result with the previous ones can be checked by 
measuring the shape of the FoF halos that are identified in our halo finding
before the step identifying the PSB halos. 
Figure 11 shows the relations between $c/a$ and halo mass, and those
between $\lambda$ and $c/a$ for 
FoF (dots and filled circles), isolated (open circles), central (boxes), 
and satellite (open triangles) halos.
The FoF halos are most spherical when their mass is near 
$2\times 10^{12}h^{-1} {\rm M}_{\odot}$. 
Note that $c/a$ has no maximum for isolated halos and is increasing as 
the halo mass increases. The spin is a monotonically decreasing
function of $c/a$ for all types of halo. Figure 11 agrees very well
with Figure 12 and 13 of Bett et al. (2007).

The overall message of Figure 10 seems that the internal physical properties
of dark halos are mainly controlled by small-scale physics. This is supported
by the fact that halo properties depend most sensitively on the local density
rather than on the gravitational potential.
It is also consistent with the finding of Park \& Choi (2009) and Park \&
Hwang (2009) that galaxy properties in the general field and clusters 
are significantly affected by gravitational and hydrodynamic interactions with the nearest
neighbor galaxy.
However, as Figure 8 shows, the gravitational potential has
information independent of that of the local density, and it is
worth to explore if the parameters like the gravitational potential or local shear
can tell something more about galaxy formation.

\section{Summary and Discussion}

In this paper we demonstrated the gravitational potential and its functions
can be reasonably accurately estimated from an observational sample
that covers only a finite volume of the universe and suffers from tracer biasing
and redshift space distortion.
We found that the error in the gravitational potential and shear tensor 
decreases rapidly as one moves inside the survey boundaries. In the case of 
shear tensor the error becomes less than 20\% of its RMS value in the regions
separated from the survey boundaries by more than about 30$h^{-1}$Mpc.
This requires the sample size to be much larger than 60$h^{-1}$Mpc for
an environment study with accurate estimation of the potential and its functions.
Our study also shows that the effects of halo biasing on the gravitational 
potential estimation can be greatly reduced by weighting dark halos (or galaxies)
by their mass as was done by Park et al. (2008) and Park \& Choi (2009) in their
studies of small and large-scale density dependence of galaxy properties. 
Accuracy in the estimation starts to fall down rapidly when the halo 
mass cut is larger than $10^{13}h^{-1}$M$_{\odot}$i. This means that
the mass density and potential fields estimated from the distribution of
the Luminous Red Galaxies (LRGs) will be quite inaccurate because
the mean separation of a volume-limited sample of the SDSS LRGs
is about 20$h^{-1}$Mpc (Gott et al. 2009), which corresponds to the 
halo mass cut of about $2\times 10^13 h^{-1} M_{\odot}$.

The error due to the redshift space distortion effects can be also reduced by using
dark halo mass density in the estimation of potential. But even more reduction
can be achieved by correcting the observed (i.e. redshift space) distribution 
of dark halos for the peculiar velocity. It was sufficient to use the peculiar
velocity linearly estimated from the redshift space distribution of dark halos.
After making the peculiar velocity correction and using halo mass weight
the error due to the redshift space distortion becomes tiny.

We showed there exists large dispersion in the gravitational potential and the shear
at fixed local density. It demonstrates the potential
has large-scale information independent of local density. 
We inspected the dependence of dark matter halo properties on
local density, gravitational potential, shear ellipticity and prolateness. 
Among these environmental parameters the local density shows the strongest
correlation with the internal physical parameters of dark halos. When halo mass
is fixed, the spin and shape parameters are nearly independent of the potential
and shear tensor but depend sensitively on the background density in the case
of massive halos, in particular.

In the following paper we will analyze the Main Galaxy sample of the SDSS DR7 catalog
to examine the dependence of various galaxy properties on these environmental 
parameters. This will extend our understanding on the environmental effects
on galaxy formation and evolution.

\acknowledgments
The authors thank the anonymous referee for valuable comments.
HBP thanks Prof. Hyung-Mok Lee for useful comments.
JHK appreciates the support of the College of Applied Science, 
Kyung Hee University research professor fellowship in CAS-RP-KJH2009.
CBP acknowledges the support of the Korea Science and Engineering
Foundation (KOSEF) through the Astrophysical Research Center for the
Structure and Evolution of the Cosmos (ARCSEC).

{}
\end{document}